\documentclass[reprint,superscriptaddress,amssymb,amsmath,aip,apl,longbibliography]{revtex4-2}

\usepackage{floatrow}
\usepackage{graphicx}
\usepackage{amssymb}
\usepackage{epstopdf}
\usepackage{upgreek}
\usepackage{dcolumn}
\usepackage{bm}
\usepackage{gensymb}
\usepackage{braket}
\usepackage{amsmath}
\usepackage{mathtools}
\usepackage{float}
\usepackage{tikz}
\usepackage[colorlinks=true,allcolors=blue]{hyperref}

\usepackage{array}

\usepackage{xcolor}

\usepackage[normalem]{ulem}

\usepackage[resetlabels,labeled]{multibib}

\expandafter\ifx\csname package@font\endcsname\relax\else
\expandafter\expandafter
\expandafter\usepackage
\expandafter\expandafter
\expandafter{\csname package@font\endcsname}%
\fi

\newcommand{\micro}{$\upmu$}
\newcommand{\microamp}{$\upmu$A}

\newcommand{\be}{\begin{eqnarray}}
\newcommand{\ee}{\end{eqnarray}}
\newcommand{\bfig}{\begin{figure}}
	\newcommand{\efig}{\end{figure}}

\newcommand{\umux}{\micro mux}
\newcommand{\microphinotrthz}{\micro$\Phi_0/\sqrt{\mathrm{Hz}}$}

\DeclareFontFamily{U}{mathb}{}
\DeclareFontShape{U}{mathb}{m}{n}{
	<-5.5> mathb5
	<5.5-6.5> mathb6
	<6.5-7.5> mathb7
	<7.5-8.5> mathb8
	<8.5-9.5> mathb9
	<9.5-11.5> mathb10
	<11.5-> mathbb12
}{}

\begin{document}
	
	\title{Improved microwave SQUID multiplexer readout using a kinetic-inductance traveling-wave parametric amplifier}
	\author{M. Malnou}
	\email{maxime.malnou@nist.gov}
	\affiliation{National Institute of Standards and Technology, Boulder, Colorado 80305, USA}
	\affiliation{Department of Physics, University of Colorado, Boulder, Colorado 80309, USA}
	\author{J. A. B. Mates}
	\affiliation{National Institute of Standards and Technology, Boulder, Colorado 80305, USA}
	\author{M. R. Vissers}
	\affiliation{National Institute of Standards and Technology, Boulder, Colorado 80305, USA}
	\author{L. R. Vale}
	\affiliation{National Institute of Standards and Technology, Boulder, Colorado 80305, USA}	
    \author{D. R. Schmidt}
	\affiliation{National Institute of Standards and Technology, Boulder, Colorado 80305, USA}    
    \author{D. A. Bennett}
	\affiliation{National Institute of Standards and Technology, Boulder, Colorado 80305, USA}	
	\author{J. Gao}
	\affiliation{National Institute of Standards and Technology, Boulder, Colorado 80305, USA}
	\affiliation{Department of Physics, University of Colorado, Boulder, Colorado 80309, USA}
    \author{J. N. Ullom}
	\affiliation{National Institute of Standards and Technology, Boulder, Colorado 80305, USA}
	\affiliation{Department of Physics, University of Colorado, Boulder, Colorado 80309, USA}
	\date{\today}
	
	\begin{abstract}    
    We report on the use of a kinetic-inductance traveling-wave parametric amplifier (KITWPA) as the first amplifier in the readout chain of a microwave superconducting quantum interference device (SQUID) multiplexer (\umux). This \umux{} is designed to multiplex signals from arrays of low temperature detectors such as superconducting transition-edge sensor microcalorimeters. When modulated with a periodic flux-ramp to linearize the SQUID response, the flux noise improves, on average, from $1.6$\,\microphinotrthz{} with the KITWPA off, to $0.77$\,\microphinotrthz{} with the KITWPA on. When statically biasing the \umux{} to the maximally flux-sensitive point, the flux noise drops from $0.45$\,\microphinotrthz{} to $0.2$\,\microphinotrthz. We validate this new readout scheme by coupling a transition-edge sensor microcalorimeter to the \umux{} and detecting background radiation. The combination of \umux{} and KITWPA provides a variety of new capabilities including improved detector sensitivity and more efficient bandwidth utilization.
    
	\end{abstract}

	\maketitle
 
    Over the past few years, the multiplexed readout of transition-edge sensors (TES) with microwave superconducting quantum interference device (SQUID) multiplexers (\umux) has become ubiquitous. For example, this multiplexing technique will be deployed at the Simons Observatory to read out signals from tens of thousands of TES bolometers, aimed at measuring the cosmic microwave background \cite{McCarrick2021The}. It is also being used to read out TES microcalorimeter arrays for x-ray and gamma-ray spectroscopy \cite{Noroozian2013high,mates2017simultaneous,Yoon2018toward,Nakashima2020low,Carpenter2021hyperspectral,Szypryt2021design,szypryt2022tabletop}.  





    
    A \umux{} divides the available readout bandwidth by coupling many readout resonators to a single transmission line. Each resonator is terminated by an rf-SQUID, inductively coupled to a TES, and this coupling is typically made large enough to ensure that the TES current noise dominates over the noise of the readout chain. For pulsed TES signals, this large inductive coupling results in a high flux slew rate at the SQUIDs, requiring a wide resonator bandwidth to track\cite{supplementary}. In this context, having a lower readout noise would allow for a smaller coupling, a slower slew rate, narrower resonators, and therefore would allow us to increase the multiplexing factor.

    
    
    Other sensors could benefit from a lower readout noise, in particular metallic magnetic calorimeters (MMCs). These devices place a magnetically susceptible calorimeter in the field of a superconducting loop that also passes through the input coil of a SQUID. With magnetic flux trapped in the loop, a variation in the MMC susceptibility shifts a fraction of this flux into or out of the SQUID. This shift of magnetic energy cannot be better resolved by increasing the inductive coupling to the SQUID. Furthermore, existing \umux{} readout techniques substantially degrade the performance of MMCs \cite{wegner2018microwave}. The use of a near-quantum-limited microwave amplifier could mitigate this problem and advance the use of multiplexed MMC arrays.

    Traditionally, the first amplifier in the \umux{} readout chain is a high electron-mobility transistor (HEMT) amplifier, placed at 4 kelvin \cite{Noroozian2013high,mates2017simultaneous,dober2017Microwave,dober2021a}. The HEMT offers several key features that make it compatible with \umux{} readout: (i) it provides sufficient gain, (ii) it is wideband, and (iii) it has a high compression power. However, its noise temperature, usually a few kelvin, is far from the lower bound imposed by quantum mechanics \cite{caves1982quantum}. 
    
    In this letter, we use a kinetic-inductance traveling-wave parametric amplifier (KITWPA) as the first amplifier in the readout chain of a \umux{}. The KITWPA gain, bandwidth and power handling are also compatible with \umux{} readout, and its wideband noise has been shown to be close to the quantum limit \cite{malnou2021three,malnou2022performance}. It is placed before the HEMT, at millikelvin temperatures. With this readout scheme, we show that the flux noise attached to a coherent tone probing one of the resonators in the \umux{}, whose resonance is modulated at $3$\,MHz, is, on average, $0.77$\,\microphinotrthz. When the KITWPA is turned off, the flux noise obtained with the HEMT as the first amplifier is more than doubled, reaching $1.6$\,\microphinotrthz{}, a typical noise level for traditional \umux{} readout chains \cite{Bennett2014Integration,dober2021a}. When the resonator is biased at its maximum flux-sensitive point, the open-loop flux noise drops from $0.45$\,\microphinotrthz{} (with the KITWPA off) to $0.2$\,\microphinotrthz{} (with the KITWPA on), equivalent to a system noise temperature of about $1$\,K. Such a low open-loop flux noise suggests that the demodulated flux noise could be decreased even further with straightforward improvements. Finally, we validate the use of the KITWPA coupled to the \umux{} by measuring pulse events in a TES microcalorimeter caused by background radiation.
    
\begin{figure}[h!]
	\centering
	\includegraphics[scale=0.35]{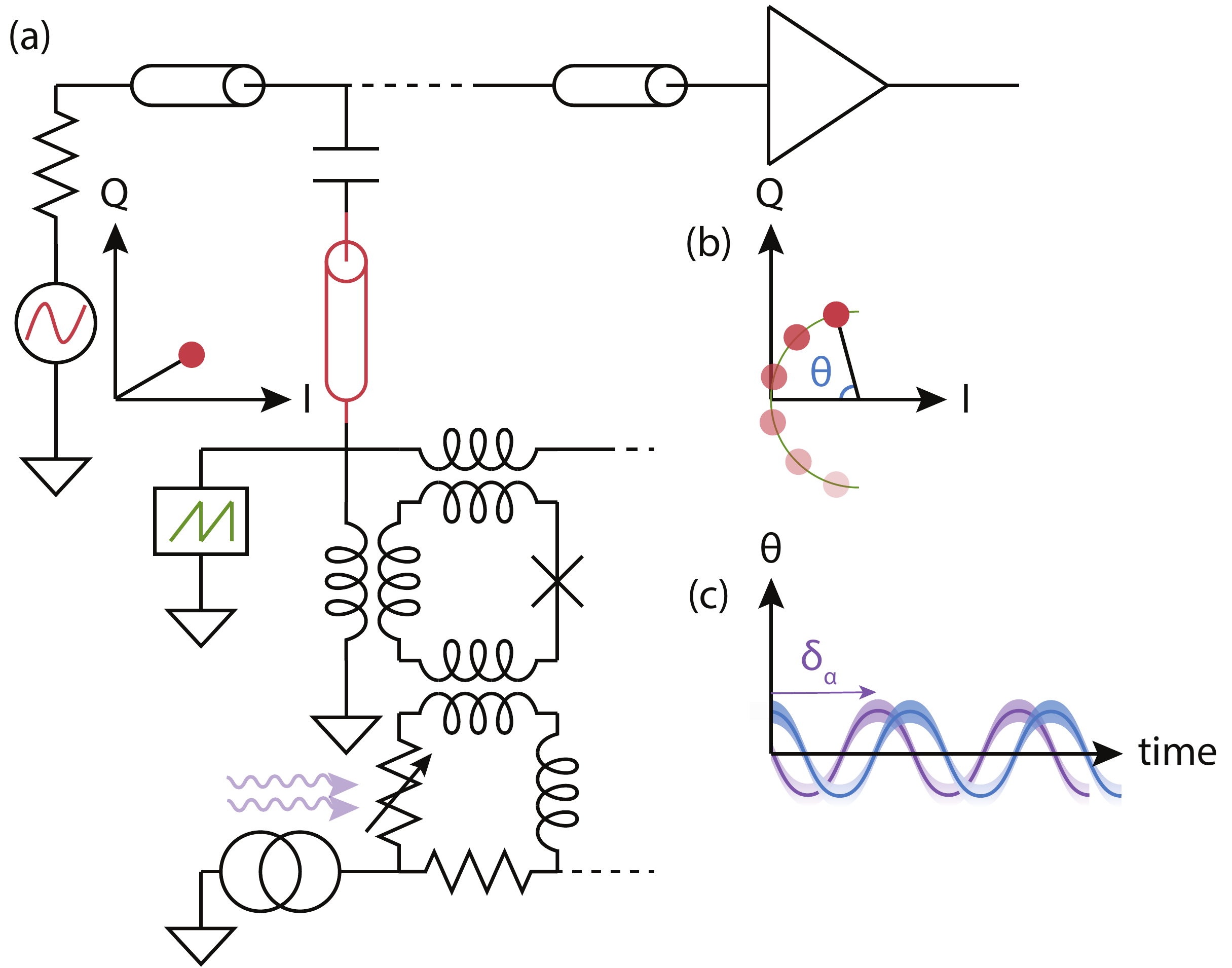} 
	\caption{Schematic of a \umux{} readout circuit, showing how amplifier noise translates to flux noise. In traditional \umux{} readout (a) a TES (shown as a variable resistor) is inductively coupled to a radio frequency (rf) SQUID embedded in a microwave resonator. (b) For a fixed microwave probe tone, the transmission moves along the resonance circle as a periodic function of flux in the SQUID. The noise associated with the amplifier chain (red disk) spreads the tone's position along the I and Q quadratures. (c) Under flux-ramp modulation, the SQUID is constantly sweeping out its approximately sinusoidal response. The noise attached to the tone thus translates into some noise on $\theta(t)$, the phase of the flux-ramp response. This noise impacts how well one can detect a phase shift $\delta_a$ on $\theta(t)$, due to a TES signal.}
    \label{fig:umux_schem}
    \end{figure}

    
    Qualitatively, for microwave SQUID multiplexing, a low amplifier noise translates into a low flux noise. Each resonator within a \umux{} is terminated by an rf-SQUID, whose loop is coupled to a flux-bias line used to modulate the resonator's resonant frequency (see Fig.\,\ref{fig:umux_schem}a). A probe tone, with frequency centered within the peak-to-peak frequency shift of the resonance, then describes a semi-circle trajectory in its rotating frame, at the modulation frequency. The readout noise attached to this tone can then be thought of as spreading its instantaneous position over a two-dimensional Gaussian in the rotating frame, along the in-phase (I) and out-of-phase (Q) quadratures (see Fig.\,\ref{fig:umux_schem}b). It determines the noise on the tone's angle $\theta$ in this frame, and in turn, it determines the noise on the phase of $\theta$ (see Fig.\,\ref{fig:umux_schem}c). This phase noise, multiplied by $\Phi_{0}/2\pi$, gives the flux noise.
    
    
    To reduce the flux noise, one can (i) increase the tone's power, but this power is eventually limited by the linearity of the SQUID response, and (ii) reduce the noise of the readout chain, which is what we propose to do, using a near-quantum-limited amplifier. In this context, for a given tone's power we can ask: what is the lowest flux noise achievable? In other words, what is the flux noise associated with a quantum-limited amplification chain? 

    
    Quantitatively now, starting with a system noise temperature $T_\mathrm{sys}$, the (input-referred) noise power spectral density $S_N$ along the I and Q quadratures of the tone is $S_N=k_BT_\mathrm{sys}$, where $k_B$ is the Boltzmann constant. Normalizing by the tone's power $P_t$, it translates into a spectral density on the rotation angle of the tone, $\theta$, such that $S_\theta=4 k_B T_\mathrm{sys}/P_t$ \cite{noisenote}. Assuming a sinusoidal variation of $\theta$ with the flux $\Phi$, $\theta(\Phi)=A\cos(2\pi\Phi/\Phi_0)$, with $A$ the variation amplitude and $\Phi_0$ the magnetic flux quantum, the maximum slope is then $\max\{d\theta/d\Phi\}=2\pi A/\Phi_0$. At this maximum flux-sensitive point, the noise power spectral density on the flux is $\Tilde{S}_\Phi=S_\theta\Phi_0^2/(2\pi A)^2$. Therefore, the flux noise $\sqrt{\Tilde{S}_\Phi}$ at the maximum flux-sensitive point, sometimes called the \textit{open-loop} flux noise, is:
    \begin{equation}
    \label{eq:sigmatilde}
        \sqrt{\Tilde{S}_\Phi}=\frac{1}{\pi A}\sqrt{\frac{k_B T_\mathrm{sys}}{P_t}}.
    \end{equation}
    For typical \umux{} operation, $A\simeq1$; thus, with a system noise temperature $T_\mathrm{sys}=4$\,K, representative of a HEMT, and a probe tone power $P_t=-75$\,dBm, we obtain $\sqrt{\Tilde{S}_\Phi}=0.42$\,\microphinotrthz. Note that the dependence of the flux noise on the system noise temperature enters as a square root (due to the conversion from a power noise to an amplitude noise), so in practice, significantly reducing the flux noise is a difficult task. Note also that Eq.\,\ref{eq:sigmatilde} gives the most honest way to quote $T_\mathrm{sys}$ knowing $\sqrt{\Tilde{S}_\Phi}$, because here the system noise temperature includes all the possible sources of noise that contribute to the flux noise.

    At the standard quantum limit (SQL), the noise power spectral density is equal to one photon \cite{caves1982quantum}, $S_N=\hbar\omega$, where $\hbar$ is the reduced Planck constant and $\omega$ is the angular frequency of the photon. Thus, at the SQL, $\sqrt{\Tilde{S}_\Phi^\mathrm{SQL}}=1/(A\pi)\sqrt{\hbar\omega/P_t}$. For $\omega=2\pi\times4.5$\,GHz and with  $P_t=-75$\,dBm (and $A=1)$ it means that $\sqrt{\Tilde{S}_\Phi^\mathrm{SQL}}=0.1$\,\microphinotrthz. This is the quantum limit on the flux noise at the maximum flux-sensitive point, for this photon frequency and probe tone power.

    
    In practice, the SQUIDs within the \umux{} are always modulated with a fast ramp of current\cite{Mates2012flux}, transforming the detector signal into a phase shift of the SQUID modulated flux response. This technique evades multiple sources of low-frequency noise. In this context, the flux noise $\sqrt{S_\Phi}$ on the demodulated tone is obtained from the open-loop flux noise, degraded twice: (i) when the peak-to-peak frequency shift of the resonator $\delta_f$ is comparable to its bandwidth B and when the modulation of $\theta$ is sinusoidal, the flux power spectral density integrated over the full $2\pi$ modulation of $\theta$ increases\cite{Mates2012flux} by a factor of $\approx2$. (ii) Usually, the beginning of each ramp of current has a transient, which affects the modulation of $\theta$. The transient is eliminated by discarding the first $\Phi_0$ of the ramp response. Thus, if the ramp is swept over $n\Phi_0$, the noise increases by a factor $1/\alpha=n/(n-1)$. The general expression of the tone's \textit{demodulated} flux noise is thus:
    \begin{equation}
    \label{eq:fluxnoise}    
    \sqrt{S_\Phi} = \frac{\sqrt{2/\alpha}}{\pi A}\sqrt{\frac{k_B T_\mathrm{sys}}{P_t}}.
    \end{equation}
    For example, with $A=1$, $\alpha=2/3$, $P_t=-75$\,dBm, and at the SQL where $k_B T_\mathrm{sys}=\hbar\omega$, with $\omega=2\pi\times4.5$\,GHz, we would predict $\sqrt{S_\Phi^\mathrm{SQL}}=0.17$\,\microphinotrthz.

    \begin{figure}[h!]
	\centering
	\includegraphics[scale=0.4]{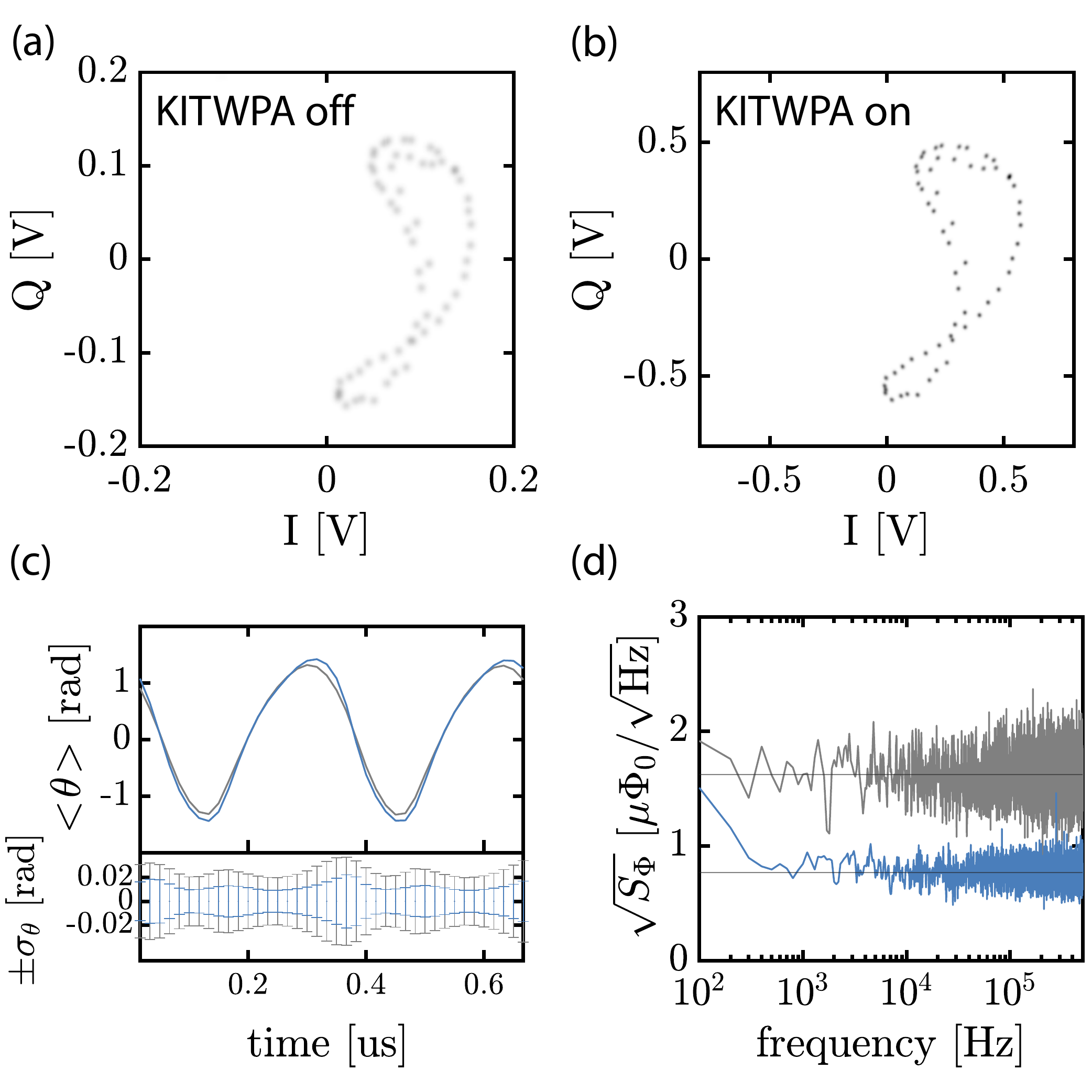} 
	\caption{Readout of a single resonator within the \umux{}, disconnected from a TES. The digitized output (at $60$\,MS/s) of the demodulated probe tone is obtained when the KITWPA is (a) off and (b) on; the origin of the I and Q quadrature frame is translated to the center of the circle supporting the tone's positions. The flux ramp sweeps 3$\Phi_0$, at a frequency $f_r=1$\,MHz (for 60 samples per ramp, 20 samples per $\Phi_0$). The tone's power is $P_t\simeq-75$\,dBm, and the tone's frequency is $f_t=4.383$\,GHz, to be compared to the resonator's maximal resonance $f_0=4.392$\,GHz and peak-to-peak frequency shift $\delta_f=15$\,MHz. (c) For each sample within the ramp (excluding the 20 first samples corresponding to the first $\Phi_0$) we measure a mean value and standard deviation for $\theta$, the angle of rotation of the tone in the I and Q quadrature frame, when the KITWPA is off (gray line) and on (blue line). (d) Extracting a value for the phase of $\theta$ for each ramp segment, we Fourier transform this phase vector (and divide by $2\pi$) to obtain the flux noise for the two situations: KITWPA off (gray line) and KITWPA on (blue line).}
    \label{fig:flux_noise}
    \end{figure}
    
    Microwave loss and excess noise will prevent the flux noise from reaching the SQL, so to see how much we can improve the flux noise in practice, we perform the readout of a \umux{}, using a KITWPA as our first amplifier, placed at millikelvin temperatures. Figure \ref{fig:flux_noise} shows the results of a single resonator readout within the \umux{}, for two situations: when the KITWPA is turned off (the HEMT is then the first amplifier in the chain), and when the KITWPA is turned on. This resonator is not connected to any TES, because otherwise the TES noise would overwhelm the readout noise (due to the engineered SQUID inductive coupling to the TES). We send a probe tone and apply a $1$\,MHz flux ramp to the SQUID, sweeping 3$\Phi_0$ per ramp so that the resonance is modulated at $f_m=3$\,MHz. Figure \ref{fig:flux_noise}a (Fig.\,\ref{fig:flux_noise}b) shows a histogram of the digitized output of the tone in its I and Q quadrature frame, obtained using a homodyne setup \cite{supplementary}, when the KITWPA is off (on). The tone frequency and power are adjusted in this frame: at the optimal tone frequency the transmission describes a ``figure-8'' shape, due to the fact that the resonator is constantly driven out of equilibrium, and above the optimal probe tone power, $P_t\simeq-75$\,dBm, the resonator bifurcates. Qualitatively, when the KITWPA is turned on, the successive positions taken by the probe tone along the flux ramp are better defined, indicative of a lower phase noise. Quantitatively, Figure \ref{fig:flux_noise}c shows $\langle\theta\rangle$ and $\sigma_\theta$, respectively the mean value and standard deviation of $\theta$ over one ramp period (discarding the transient-contaminated first $\Phi_0$) when the KITWPA is off and on. Clearly, turning on the KITWPA reduces $\sigma_\theta$. It translates into a lower flux noise \cite{supplementary}, $\sqrt{S_\Phi}$ (Fig.\,\ref{fig:flux_noise}d), which drops from $1.6$\,\microphinotrthz{} to $0.77$\,\microphinotrthz{} on average.

        
    Here, we compare the flux noise obtained when the KITWPA is off to when it is on, but the KITWPA-off situation is not equivalent to a standard \umux{} readout chain, with the HEMT as the first amplifier, because our chain contains extra microwave components \cite{supplementary}. These components insert loss, and therefore increase the flux noise. Nonetheless, even with these components the flux noise with the KITWPA off remains low compared to standard values \cite{Bennett2014Integration,dober2021a}. Furthermore, we used off-the-shelf microwave components to build the readout chain. These could be made more efficient, for example by integrating them on-chip, together with the KITWPA.

    \begin{figure}[h!]
	\centering
	\includegraphics[scale=0.4]{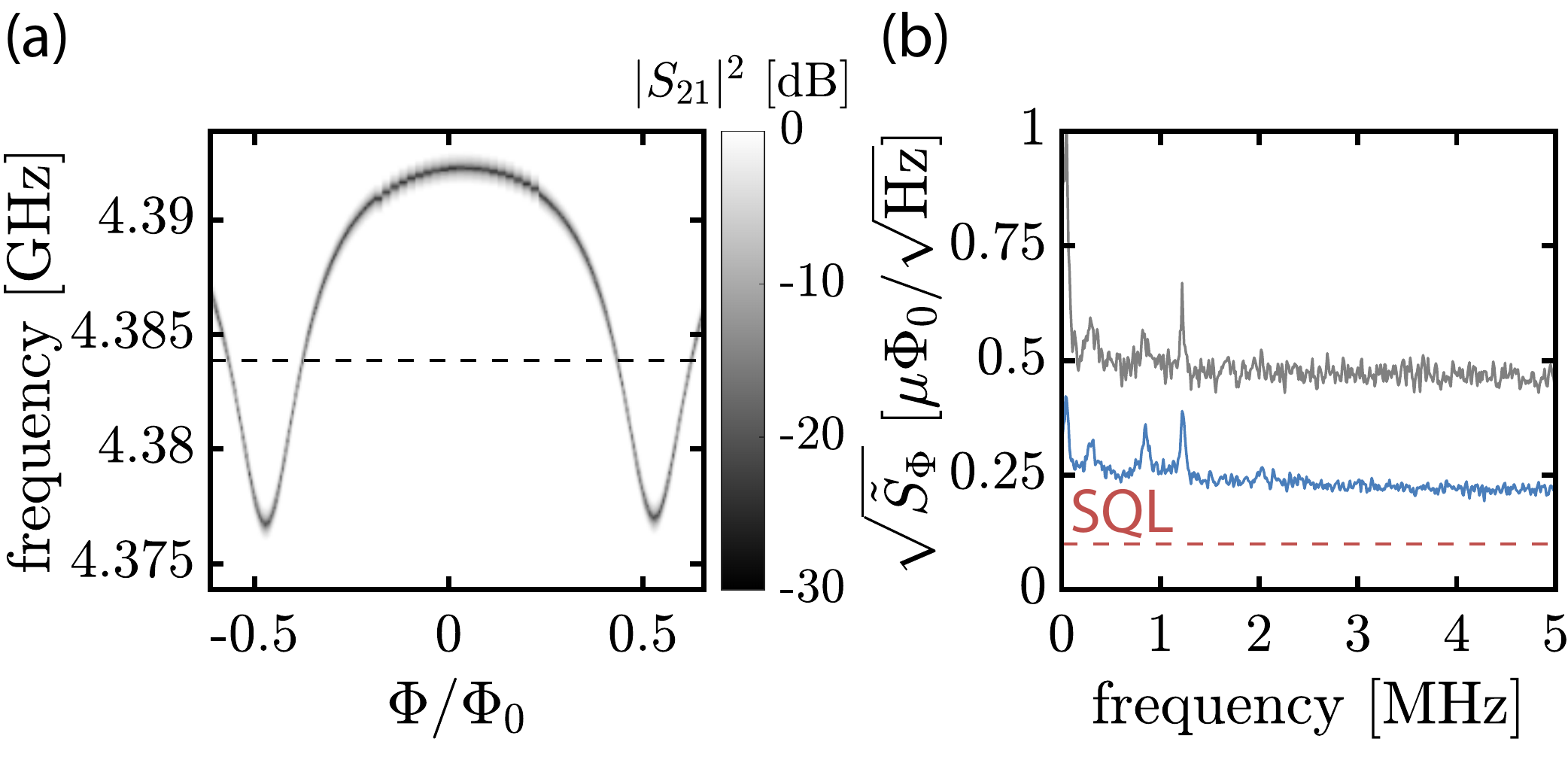} 
	\caption{Open-loop flux noise measurement. (a) The flux modulation curve of the resonator shows that $\delta_f=15$\,MHz, while the bandwidth of the resonator is $B=4.6$\,MHz. We set the tone's frequency $f_t$ at $\max\{d\theta/d\Phi\}$ (dashed line), where $f_t=4.384$\,GHz. (b) We measure the flux noise at this particular flux bias point \cite{supplementary}, when the KITWPA is off (gray line) and on (blue line). In comparison, the flux noise of an amplification chain operating at the SQL is indicated by the dashed red line.}
    \label{fig:open_loop_flux_noise}
    \end{figure}
    
    How close are we to the SQL? We cannot directly derive $T_\mathrm{sys}$ from Eq.\,\ref{eq:fluxnoise}, because in our case $\delta_f\gg B$, see Fig.\,\ref{fig:open_loop_flux_noise}a. It degrades $\sqrt{S_\Phi}$ compared to the situation where $\delta_f=B$, because when sweeping the ramp the tone then spends more time away from resonance, in the flux-insensitive region. Instead, to estimate $T_\mathrm{sys}$, we measure the open-loop flux noise at $\max\{d\theta/d\Phi\}$ \cite{supplementary}, for the two situations, KITWPA off and on, see Fig.\,\ref{fig:open_loop_flux_noise}b. At 3\,MHz, equal to the modulation frequency $f_m$ previously used, $\sqrt{\Tilde{S}_\Phi}$ drops from $0.45$\,\microphinotrthz{} to $0.2$\,\microphinotrthz{}. Indeed, the open-loop flux noise is degraded by more than $\sqrt{2/\alpha}$ (with $\alpha=2/3$) to yield the demodulated flux noise previously obtained. Using Eq.\,\ref{eq:sigmatilde}, $\sqrt{\Tilde{S}_\Phi}$ corresponds to a system noise temperature $T_\mathrm{sys}=4.6$\,K and $T_\mathrm{sys}=0.9$\,K, respectively (taking $A=1$, and $P_t=-75$\,dBm). In comparison, at the SQL, $T_\mathrm{sys}^\mathrm{SQL}=0.2$\,K (at 4.5\,GHz), therefore with the KITWPA turned on we operate $4.5$ times above the quantum limit, whereas with the KITWPA off we operate more than $20$ times above the quantum limit.
    
    Reaching such low values for $\sqrt{\Tilde{S}_\Phi}$ suggests two ways $\sqrt{S_\Phi}$ could be further reduced: (i) a resonator for which $\delta_f=B$ should, in principle, yield $\sqrt{S_\Phi}=\sqrt{2/\alpha}\sqrt{\Tilde{S}_\Phi}$, so with $\alpha=2/3$, $\sqrt{S_\Phi}$ could be as low as $0.35$\,\microphinotrthz{} (with the KITWPA on). (ii) With a tone tracking technique, where the tone's frequency is also modulated to follow the resonance, having $\delta_f\gg B$ becomes beneficial; in principle, this technique allows for a lower demodulated flux noise than the one obtained with a fixed-frequency tone \cite{yu2023slac}.
    
\begin{figure}[h!]
	\centering
	\includegraphics[scale=0.4]{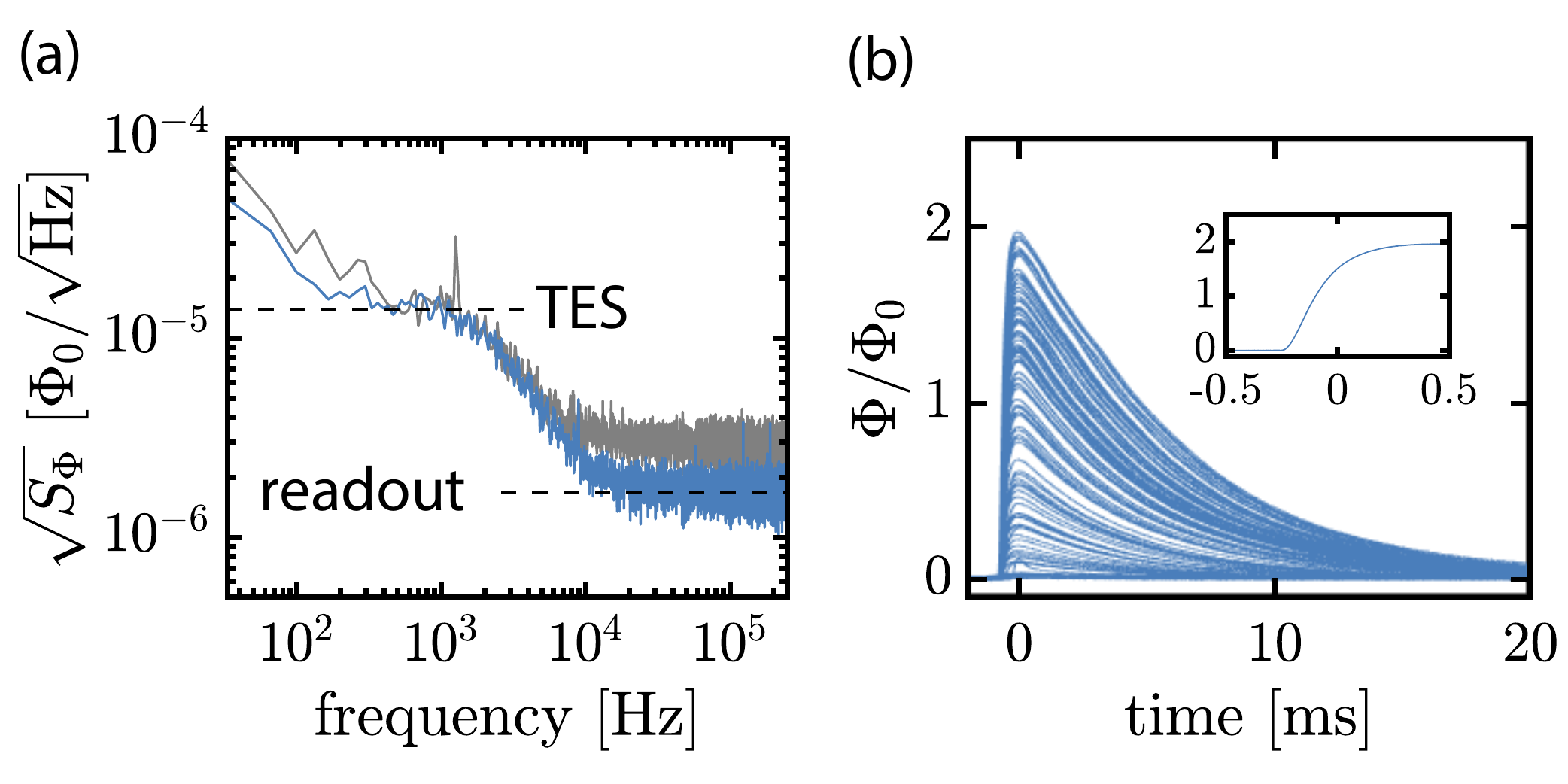} 
	\caption{Response of a resonator within the \umux{}, connected to a TES. (a) When the TES is biased between its normal and superconducting branches, the flux noise $\sqrt{S_\Phi}$ increases below 10\,kHz, because the TES noise overwhelms the readout noise. At higher frequencies, we recover the improvement in flux noise when turning the KITWPA on (blue curve) compared to when the KITWPA is off (gray curve). Note that the flux noise values are higher here than in Fig.\,\ref{fig:flux_noise}d, probably because of the unoptimized link (that includes long wire-bonds) between the TES and the \umux{} \cite{supplementary}. (b) Pulse events due to background radiation have been detected by the TES over 3.5 hours.  Focusing on times around 0 ms (inset) and on a single large amplitude pulse, it is evident that the KITWPA readout chain records the pulses without distortion even where the derivative of the flux signal is largest.}
    \label{fig:cosmicrays}
    \end{figure}

    To demonstrate that this amplification chain can truly be used for sensor readout, we connected another \umux{} channel to a TES\cite{supplementary}. When the TES detects a photon or a particle, it generates a pulse of current in the SQUID loop, which translates into a dephasing event on the probe tone's trajectory (see Fig.\,\ref{fig:umux_schem}). Absent any radiation, the flux noise is dominated by the TES noise at frequencies below 10\,kHz, see Fig.\,\ref{fig:cosmicrays}a, and therefore there is no difference between the two situations, KITWPA off and on. In fact, the coupling to the TES within this \umux{} has been engineered to overwhelm higher readout noises than those obtained when using the KITWPA. With a redesign of the \umux{}, where both the coupling and the resonator bandwidth could be reduced with no penalty on the readout sensitivity, one would truly benefit from the near-quantum-limited nature of the readout chain.
    
    Continuously acquiring the tone's excursion in the quadrature frame over several hours, we record the events for which $\Phi/\Phi_0$ significantly deviates from zero. These events correspond to cosmic rays and other background radiation hitting the TES \cite{supplementary}. In Fig.\,\ref{fig:cosmicrays}b, we have overlapped the 226 events where $\Phi/\Phi_0\leq2$, detected over 3.5 hours with the KITWPA on (events with $\Phi/\Phi_0>2$ are present but excluded from the plot because the TES begins to saturate). This experiment proves that the readout scheme using the KITWPA is suitable for the detection of pulsed events from the deposition of energy quanta.
       

   In conclusion, we have demonstrated an unprecedented microwave SQUID multiplexing readout sensitivity, using a near-quantum-limited KITWPA as our first amplifier. Modulating a \umux{} resonance at $3$\,MHz with a ramp, we showed that the flux noise of a demodulated tone is, on average, $\sqrt{S_\Phi}=0.77$\,\microphinotrthz{}, and it could be significantly lowered with straightforward improvements. In the context of \umux{} readout, the true system noise temperature must be calculated from the knowledge of the flux noise and the probe tone power. Here, our open-loop flux noise of $0.2$\,\microphinotrthz{} is equivalent to a system noise temperature of 0.9\,K, or about 4.5 times above the quantum limit. Continuously monitoring over several hours the output of one resonator connected to a TES, we successfully measured the dynamic response of the sensor to background radiation, validating the use of this new amplification chain. This improvement of noise temperature should allow a doubling of \umux{} multiplexing factor for TES microcalorimeter readout, as well as enable the useful application of \umux{} to MMCs.

   We thank D. T. Becker and D. S. Swetz for useful discussions. We gratefully acknowledge support from the NIST Program on Scalable Superconducting Computing, the National Aeronautics and Space Administration (NASA) under Grant No. NNH18ZDA001N-APRA, and the Department of Energy (DOE) Accelerator and Detector Research Program under Grant No. 89243020SSC000058.



    
    %

	\vspace{0.1in}	

\begin{thebibliography}{19}
\bibitem [{noi()}]{noisenote}%
		\BibitemOpen
		\href@noop {} {}\bibinfo {note} {In fact, $S_N = R^2 S_\theta$, and when
			normalizing by the tone's power, $R=1/2$ is the radius of the semi-circle in
			Fig.\,\ref{fig:umux_schem}b.}\BibitemShut {Stop}%
\end{thebibliography}
	%

\clearpage
\widetext
\begin{center}
\textbf{\large Supplementary information: Improved microwave SQUID multiplexer readout using a kinetic-inductance traveling-wave parametric amplifier}
\end{center}
\setcounter{equation}{0}
\setcounter{figure}{0}
\setcounter{table}{0}
\makeatletter
\renewcommand{\theequation}{S\arabic{equation}}
\renewcommand{\thefigure}{S\arabic{figure}}
\renewcommand{\thetable}{S\arabic{table}}
\renewcommand{\bibnumfmt}[1]{[S#1]}
\renewcommand{\citenumfont}[1]{S#1}

    \section{Bandwidth utilization with flux ramp modulation}

    In  flux-ramp modulated SQUID readout, the flux ramp rate sets the effective sampling rate of the SQUID\cite{Mates2012flux_s}, which must be high enough that the maximum flux excursion between samples is less than $\Phi_{0}/2$. The full width at half maximum $B$ of each resonator is then sized to contain the flux ramp signal, i.e. we must have
    \begin{equation}
        B > \frac{4n}{\Phi_{0}}M_\mathrm{in}\left|\frac{dI_\mathrm{TES}}{dt}\right|_{\mathrm{max}},
    \end{equation}
    where $n$ is the number of SQUID oscillations per ramp, $M_\mathrm{in}$ is the input mutual inductance between the detector and the SQUID and $I_\mathrm{TES}$ is the detector current. Reducing the flux noise allows us to reduce $M_\mathrm{in}$ and therefore to reduce $B$.
 
    \section{Experimental setup}

\begin{figure*}[t!]
	\centering
	\includegraphics[scale=0.35]{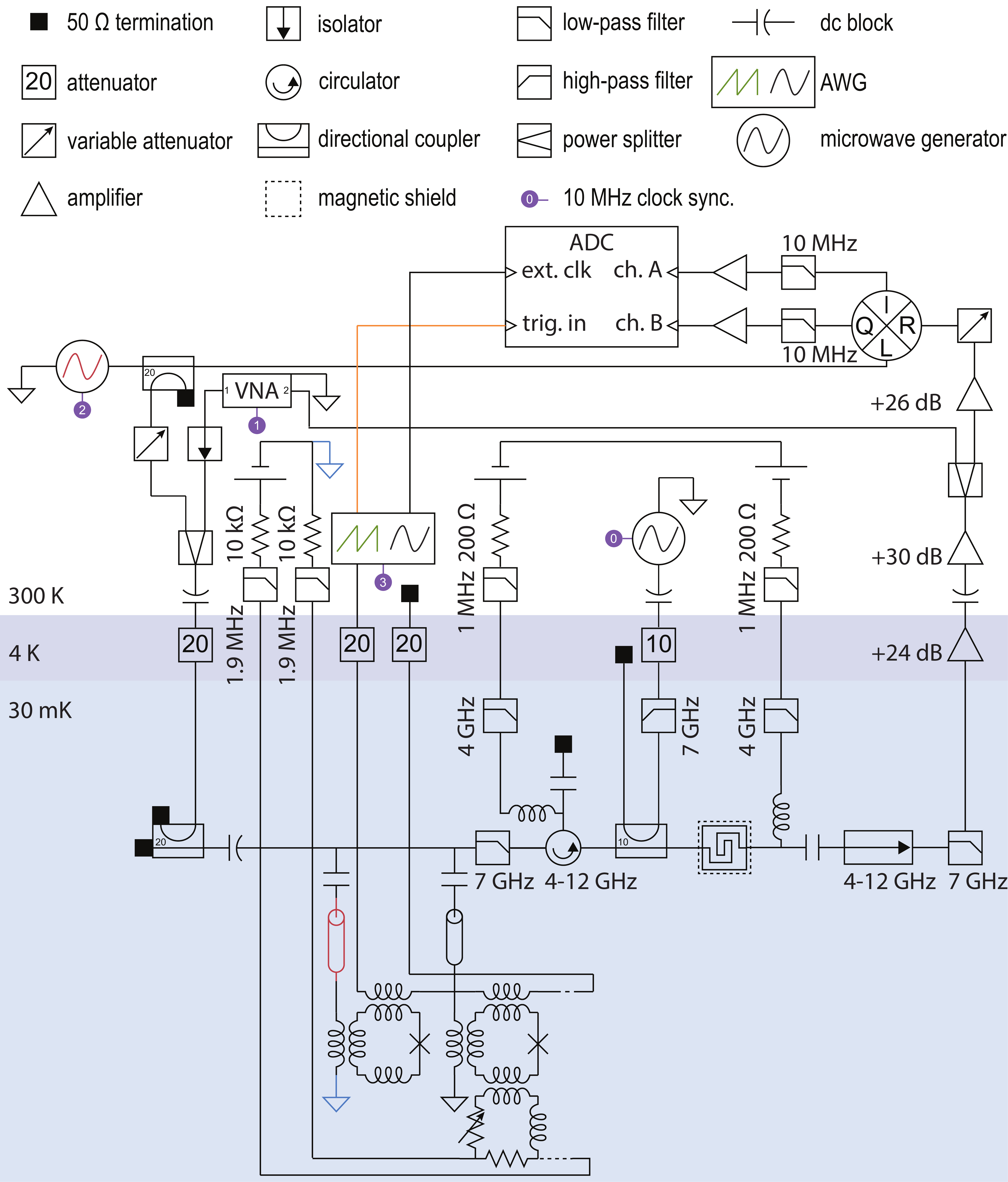} 
	\caption{Schematic of the experimental setup. We have represented only 2 of the 16 microwave resonators that the \umux{} electronic chip truly contains. In the schematic, one of the resonators is connected to a TES, while one is left unconnected. The KITWPA electronic chip is represented by the spiral inside the square. See the main text for the detailed description.}
    \label{fig:exp_setup}
    \end{figure*}

    Figure \ref{fig:exp_setup} presents a schematic of the experimental setup. At millikelvin temperatures, the input of the chain (left side) consists of a directional coupler (DC) from which we can send a vector network analyzer (VNA) tone, along with the \umux{} interrogating tone. A dc-block follows, preventing the dc current used to bias the KITWPA from leaking back toward the directional coupler and dissipating in the $50$\,\ohm{} load attached to it. The \umux{} is then connected, followed by a low-pass filter (LPF) which provides $40$\,dB rejection at the KITWPA pump frequency, preventing the strong KITWPA pump tone ($P_t\simeq -30$\,dBm) from perturbing the \umux{} resonators. In fact, even though this tone is detuned from the \umux{} resonances ($f_p=8.941$\,GHz), we have seen that it can affect the \umux{} if not properly filtered. Next, the circulator provides additional isolation, and we use its third port to send the KITWPA dc current bias, while the following DC delivers the KITWPA pump. After the KITWPA, a bias tee separates the dc bias current from the rf signal. The rf signal is then directed onto an isolator and a LPF, in order to filter out the KITWPA pump tone, which, reflected off the LPF dissipates into the isolator. The rf signal is then routed into a HEMT at 4\,K, and further amplification stages at room temperature.

    All the connecting cables going in and out of the fridge are microwave coaxial cables, even those carrying a dc signal. In fact, we have found that they provide a better isolation against radio frequency noise than twisted pair cables. For the flux ramp line, the 20\,dB attenuator at 4\,K is equivalent to a 500\,Ohm series resistor, which we use to convert the voltage excitation into a current excitation. The TES bias source is grounded to the fridge ground, but it is not connected to the electronic rack's ground. Again, the cables delivering the TES bias are microwave coaxial cables. Finally, the KITWPA dc bias source is fully floating. Care was taken not to connect the ground of the room temperature electronics to the fridge's ground through all these cables (which would create ground loops). That is why most of the microwave cables going into the fridge contain a dc block before entering the fridge. Those that don't are either fully floating, or connected to floating electronics equipment.

    At room temperature, two microwave generators are used: one for the KITWPA pump and one for the \umux{} probe tone. This second generator also serves as the local oscillator (LO) for the in-phase/quadrature (IQ) demodulator. The IQ demodulator demands a fixed input power for its LO, therefore we placed a variable attenuator on the probe tone's path. The VNA allows us to measure the KITWPA gain, and an arbitrary waveform generator (AWG) is used to generate the flux ramp. Its second channel serves as a clock for the analog to digital converter (ADC) so that the ADC can digitize at a specific rate (chosen to have an integer number of samples per flux ramp). The output trigger of the AWG is connected to the input trigger of the ADC, so that the digitizing starts at the beginning of a ramp. At the output, the amplification chain is routed to the rf-port of the IQ demodulator, and the I and Q channels are then routed (after further base-band amplification) to channels A and B of the ADC. Finally, all the microwave generators, VNA and AWG are synchronized with a 10\,MHz clock.

    Figure \ref{fig:exp_setup}a shows a close-up picture of the \umux{} in its packaging, with some of its resonators connected to a few TESs, via series inductances and shunt resistances. Obviously, this packaging is not optimized, involving long wire-bonds in between components. The input of the microwave coplanar waveguide (CPW) readout line can be seen on the left. Figure \ref{fig:exp_setup}b presents the full \umux{} and TES packaging. The flux ramp and TES biases are fed through the in-line connectors at the bottom (and through microwave coaxial cables beforehand).
    
    \begin{figure}[h!]
	\centering
	\includegraphics[scale=0.45]{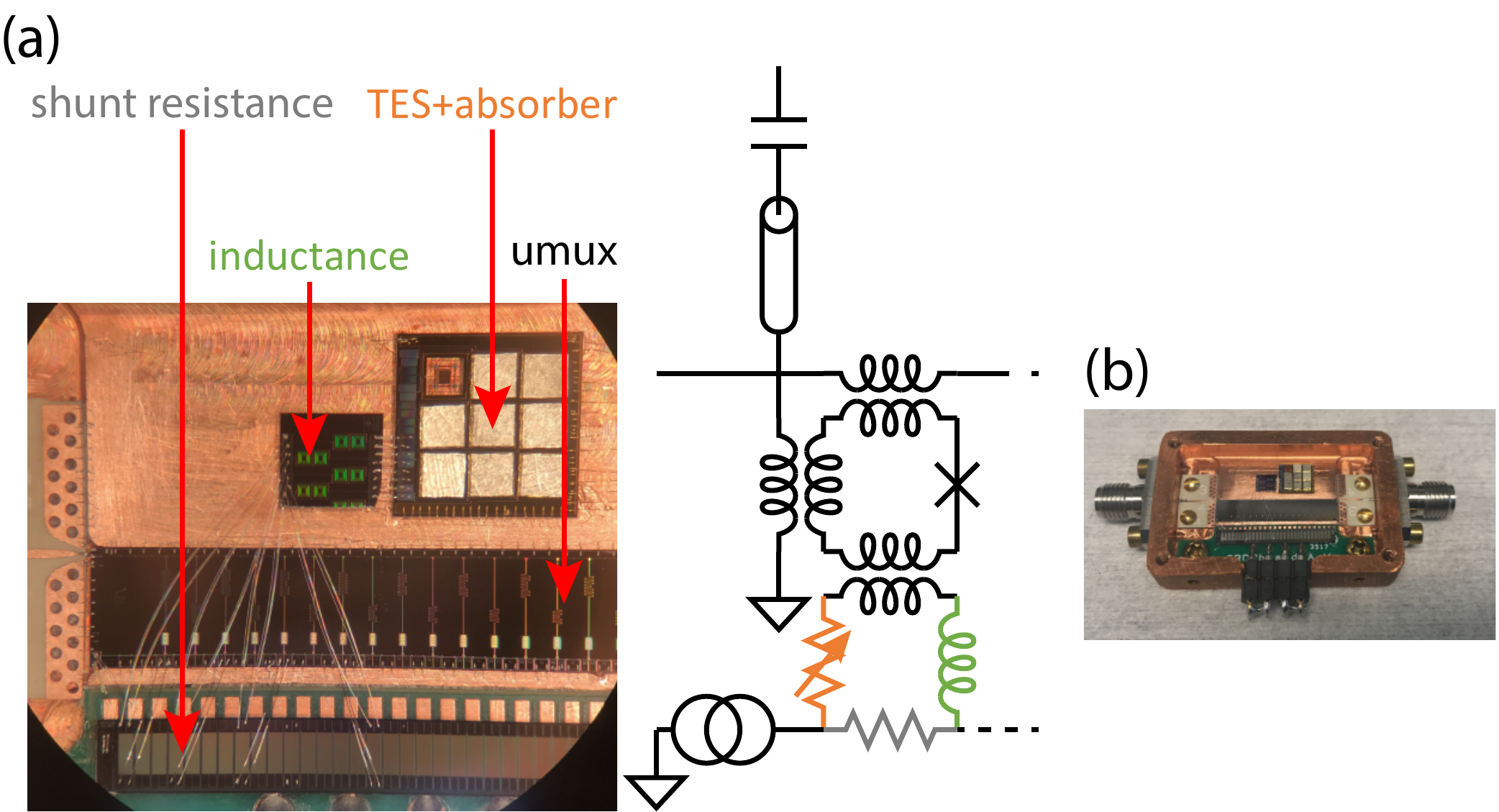} 
	\caption{Photograph of the \umux{} and TES packaging. (a) Some resonators within the \umux{} are connected to a few TES via series inductances, and shunt resistances. (b) A $1\times2"$ box contains both the \umux{} and the TES.}
    \label{fig:umux_pic}
    \end{figure}

    
    \section{Flux noise measurement, background radiation measurement and processing}

    Here we detail how, from the digitized I and Q outputs, we obtain the flux noise presented in Fig.\,\ref{fig:flux_noise}. We send a flux ramp with frequency $f_r=1$\,MHz, that sweeps $3\Phi_0$ per ramp (therefore the modulation frequency of the resonator's resonance is $f_m=3$\,MHz). We then digitize $0.1$\,s of data from the I and Q channels, at $60$\,MS/s, which means that we obtain two arrays of $6\times10^6$ voltages. It also means that we have 60 samples per ramp, or 20 samples per $\Phi_0$. We perform this acquisition twice: when the KITWPA is off (pump tone off), and when it is on.

    The processing of this data consists of several steps. First, we translate all the coordinates in the IQ rotating frame, such that the center of the frame corresponds to the center of the semi-circle trajectory described by the points. In practice, we fit all the points to a circle, and find its center. When centered, a histogram of the points positions is given by Fig.\,\ref{fig:flux_noise}a (Fig.\,\ref{fig:flux_noise}b), when the KITWPA is off (on).

    Second, we unwrap the angle $\theta$ of these points so that $\theta$ continuously varies between $-\pi$ and $\pi$ as a function of time. We then extract the phase $\alpha(t)$ from $\theta(t)$, excluding the first 20 values of $\theta(t)$ that correspond to the first $\Phi_0$, using \cite{Mates2012flux_s}:
    \begin{equation}
    \label{alpha}    
    \alpha = \arctan\left(\frac{\sum \theta(t) \sin(\omega_m t)}{\sum \theta(t) \cos(\omega_m t)}\right),
    \end{equation}
    where $\omega_m = 2\pi f_m$. The vector of phases $\alpha(t)$ runs from $t=0$\,s to $t=0.3$\,s, with a step size of $\delta_t=1/f_r$. 

    We can now calculate the power spectral density (psd) of $\alpha(t)$ to get $S_\alpha$, in units of $\mathrm{rad}^2/\mathrm{Hz}$. The maximum frequency of this psd is half the ramp frequency, and we chose a bin size of 100\,Hz. Finally, the flux noise is given by $\sqrt{S_\Phi}/\Phi_{0}=\sqrt{S_\alpha}/(2\pi)$.

    When running the experiment continuously to detect background radiation, we set the fridge temperature to $90$\,mK (stable within $\pm15$\,\micro K). We then continuously perform the acquisition and processing, and only keep the data from the I and Q channels if a significant peak in $\alpha(t)$ is detected. The procedure to look for such a peak consists of calculating the moving standard deviation $\sigma_\alpha$ of $\alpha(t)$ (with a 5\,ms window), and look for peaks in $\sigma_\alpha$ that are greater than $0.03$\, radians (a value that gives good compromise between detecting too many false positives and discarding too many low energy background radiation events).
    
    \section{Open-loop flux noise measurement and processing}

    The open-loop flux noise is acquired with the same experimental setup. However, instead of using the AWG to generate a voltage ramp, we use it as dc voltage source (which translates into a dc current, due to the 20\,dB attenuator at 4\,K). We then digitize the I and Q outputs for $221$ static flux biases along one full $\Phi_0$ (equivalent to $1.2$\,mA; we use coarse current steps of $25$ \microamp{} in the flux-insensitive region, close to the maximum resonator frequency, and fine current steps of $5$ \microamp{} in the flux-sensitive region). For each flux biasing point, and for the two configurations KITWPA off/on, we record data over $0.01$\,s, with a sampling rate of $20$\,MS/s, therefore we have two arrays of $2\times10^5$ voltages per bias point, per KITWPA configuration.

    The processing steps are as follows. First, we translate the coordinates in the IQ rotating frame, as in the case of the demodulated flux noise. For each flux bias, we find the average coordinate (two voltages) in the IQ frame, and then fit all the averages to a circle to find its center.

    Second, we calculate the voltage psd $S_V(\Phi)$ (in $V^2/\mathrm{Hz}$) for each flux bias, along the direction tangential to the circle (the unit vector in the tangential direction is the normalized gradient of the averaged points coordinates).
    
    Finally, we calculate the open-loop flux noise, as:
    \begin{equation}
        \sqrt{\Tilde{S}_\Phi} = \frac{\sqrt{S_V(\Phi_M)}}{\lvert \mathrm{max}\{dV/d\Phi\} \rvert},
    \end{equation}
    where $\Phi_M$ is the flux bias for which $\lvert dV/d\Phi \rvert = \lvert \mathrm{max}\{dV/d\Phi\} \rvert$.

    \section{KITWPA gain profile}

\begin{figure}[h!]
	\centering
	\includegraphics[scale=0.45]{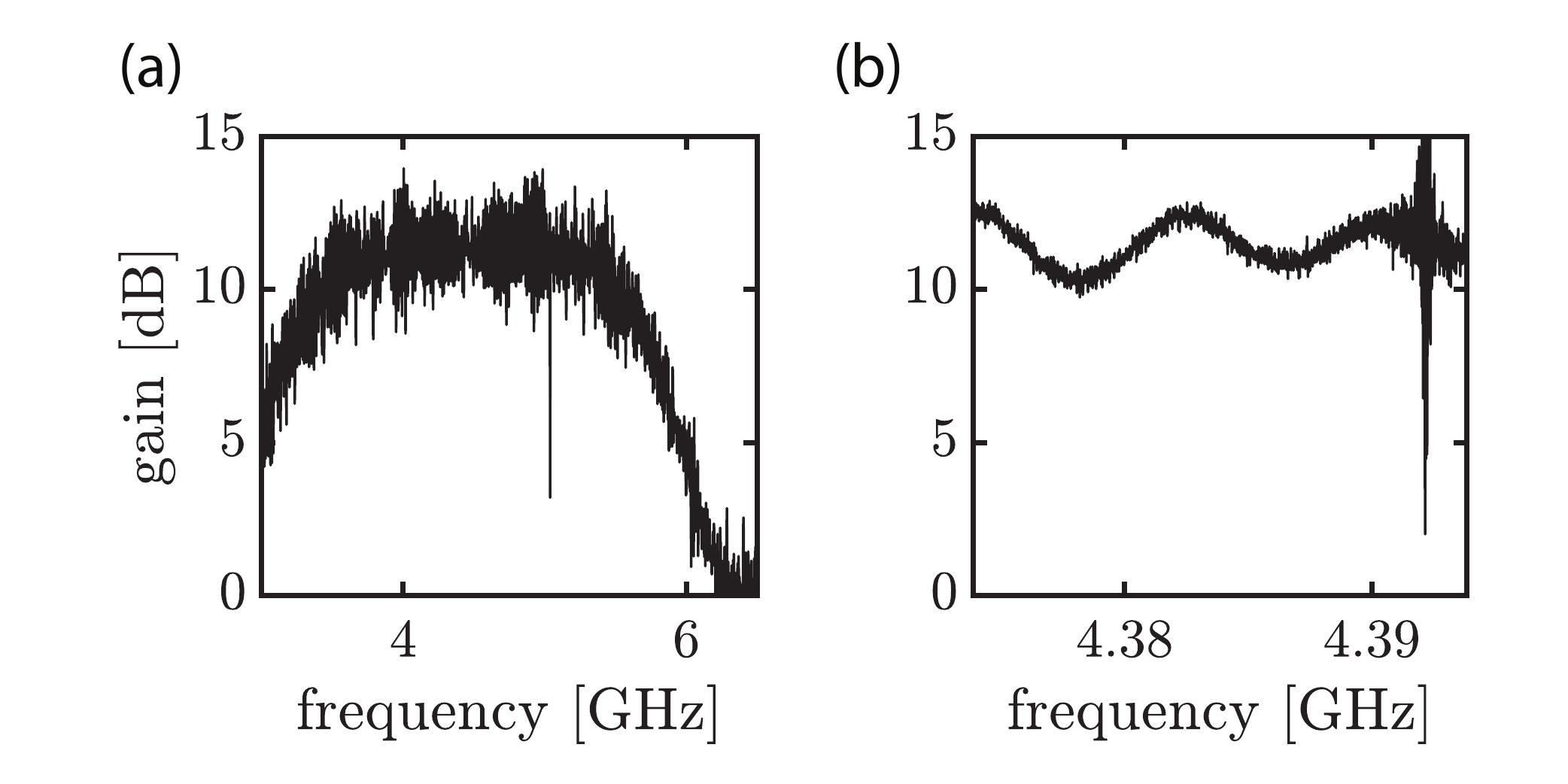} 
	\caption{Gain of the KITWPA (ratio pump on/off). (a) The wideband gain profile as a function of frequency. (b) Zoom in on the gain around the tuning range of the resonator used in the flux noise measurements.} 
    \label{fig:KITWPA_gain}
    \end{figure}

    Figure \ref{fig:KITWPA_gain}a shows the wideband KITWPA gain profile, obtained with the VNA (pump on/off ratio), at which the KITWPA was operated during the various experiments. We chose to run the KITWPA at a relatively modest gain, to ensure its performance as a function of time be very stable.
    
    Figure \ref{fig:KITWPA_gain}b shows the KITWPA gain, when zoomed in around the tuning range of the resonator that we used in the demodulated flux noise measurement. The feature at $4.392$\,GHz comes from the presence of the resonance (the VNA tone passes through the \umux{}, see fig.\,\ref{fig:exp_setup}). Over the whole resonator's tuning range, the KITWPA gain is on average $11.6$\,dB, with $2.5$\,dB of peak-to-peak gain ripples. Interestingly, because of the nature of the flux-ramp modulation scheme, the \umux{} readout is somewhat insensitive to gain ripples: it may distort $\theta(t)$ into a non perfect sinusoidal signal, but we only care about the phase of that signal. So as long as the gain ripples are stable as function of time, they don't affect the periodicity of $\theta(t)$, therefore they don't affect the determination of $\alpha(t)$.

\end{document}